\documentclass[aps,prl,reprint]{revtex4-1}

\usepackage{graphicx}% Include figure files
\usepackage{dcolumn}% Align table columns on decimal point
\usepackage{amsmath}
\usepackage{upgreek}

\begin{document}

\preprint{APS/123-QED}

\title{Simple atomic quantum memory suitable for semiconductor quantum dot single photons}

\author{Janik Wolters}
\email{janik.wolters@unibas.ch}
\affiliation{Department of Physics, University of Basel, Klingelbergstrasse 82, CH-4056 Basel,
Switzerland}

\author{Gianni Buser}
\affiliation{Department of Physics, University of Basel, Klingelbergstrasse 82, CH-4056 Basel,
Switzerland}

\author{Andrew Horsley}
\affiliation{Department of Physics, University of Basel, Klingelbergstrasse 82, CH-4056 Basel,
Switzerland}

\author{Lucas B\'eguin }
\affiliation{Department of Physics, University of Basel, Klingelbergstrasse 82, CH-4056 Basel,
Switzerland}

\author{Andreas J\"ockel}
\affiliation{Department of Physics, University of Basel, Klingelbergstrasse 82, CH-4056 Basel,
Switzerland}

\author{Jan-Philipp Jahn}
\affiliation{Department of Physics, University of Basel, Klingelbergstrasse 82, CH-4056 Basel,
Switzerland}

\author{Richard J. Warburton}
\affiliation{Department of Physics, University of Basel, Klingelbergstrasse 82, CH-4056 Basel,
Switzerland}

\author{Philipp Treutlein}
\affiliation{Department of Physics, University of Basel, Klingelbergstrasse 82, CH-4056 Basel,
Switzerland}

\date{\today}

\begin{abstract}
Quantum memories matched to single photon sources will form an important cornerstone of future quantum network technology.
We demonstrate such a memory in warm Rb vapor with on-demand storage and retrieval, based on electromagnetically induced transparency. With an acceptance bandwidth of $\delta f$ = 0.66~GHz the memory is suitable for single photons emitted by semiconductor quantum dots. 
In this regime, vapor cell memories offer an excellent compromise between storage efficiency, storage time, noise level, and experimental complexity, and atomic collisions have negligible influence on the optical coherences. 
Operation of the memory is demonstrated using attenuated laser pulses on the single photon level. For 50 ns storage time we measure $\eta_{\textrm{e2e}}^{\textrm{50ns}} = 3.4(3)\%$ \emph{end-to-end efficiency} of the fiber-coupled memory, with an \emph{total intrinsic efficiency} $\eta_{\textrm{int}} = 17(3)\%$.  
Straightforward  technological improvements can boost the end-to-end-efficiency to $\eta_{\textrm{e2e}} \approx 35\%$; beyond that increasing the optical depth and exploiting the Zeeman substructure of the atoms will allow such a memory to approach near unity efficiency.
 In the present memory, the unconditional readout noise level of $9\cdot 10^{-3}$ photons is dominated by atomic fluorescence, and for input pulses containing on average $\mu_{1}=0.27(4)$  photons the signal to noise level would be unity.
\end{abstract}





\maketitle

Quantum networks built from optical fiber-linked quantum nodes \cite{Kimble2008} open manifold opportunities across a range of scientific and technological frontiers. For example: high-speed quantum cryptography networks can be used for unconditionally secure communication in metropolitan areas \cite{Gisin2002}; and quantum networks can help realize large scale quantum computers and quantum simulators that will allow for exponential speed-up in solving complex problems \cite{Ladd2010, Kok2007}. 
Photonic quantum networks, in turn, require a scalable quantum node technology that allows for (i) storing quantum information in a quantum memory \cite{Bussieres2013}; and (ii) on-demand conversion of this information into single photons traveling along the network interconnects. 

To realize  quantum nodes, a heterogeneous approach \cite{Sangouard2007,Sangouard2011} is highly promising. Heterogeneous quantum nodes consist of a single photon source and a compatible quantum memory, where the systems may be completely different from each other and can be individually optimized. 
For the single photon source, self-assembled semiconductor quantum dots (QD) are arguably the best choice, as they allow 
for high speed on-demand photon generation with up to GHz emission rates and measured efficiencies \cite{Munsch2013, Somaschi2015, Ding2016}  as high as 75\%. 
These sources can emit indistinguishable single photons \cite{Somaschi2015, Santori2002, Ates2009}, or even polarization-entangled photon pairs \cite{Stevenson2006, Chen2016}, and the QD spin can be entangled with an emitted photon~\cite{Delteil2016,Gao2012}. However, the quantum dot itself is not a good quantum memory, since the  coherence times are limited by the comparably strong coupling to the solid-state environment.
To make this exquisite source of single or entangled photons useful for quantum networks, the QD therefore has to be combined with a  quantum memory with a high end-to-end efficiency. 
In this letter, we present such a memory  based on warm rubidium vapor. The developed memory can readily be combined with a well-engineered GaAs/AlGaAs QD single photon source \cite{Jahn2015,Munsch2013,Akopian} and may then serve as a network node, bringing the vision of functional quantum networks closer to reality.

Many different physical platforms for quantum memories are currently unter investigation, ranging from phonons in solids to atomic Bose-Einstein-condensates \cite{Heshami2015,Bussieres2013}. 
Alkali vapor cells are particularly appealing, as they require neither cryogenic temperatures, nor advanced laser cooling techniques, both of which hinder large scale or field applications. 
Ten~millisecond coherence times have been demonstrated in suitable cells~\cite{Borregaard2016}, and advances up to 100~s are possible with improved anti-relaxation coating technology~\cite{Balabas2010}, clearly sufficient for most applications.
A variety of memory protocols for alkali vapors have been developed~\cite{Lvovsky2009}, based either on absorption engineering~\cite{Hosseini2011}, or optically controlled light-matter interaction~\cite{Kupchak2015,Reim2010,Shuker2008}.  
While wavelength matching to QD photons has been achieved \cite{Akopian, Ulrich2014, Jahn2015}, a remaining challenge for building a QD compatible atomic memory is that the required acceptance bandwith of  $\delta f=$ 0.5 -- 1.0 GHz \cite{Kuhlmann2015,Kuhlmann2013} is rather large compared to the intrinsic linewidth of the alkali D lines, which is on the order of $\delta_{\textrm{Rb}}=5$~MHz \cite{Stecka}.

\begin{figure}
\center
\includegraphics[width=0.9\columnwidth]{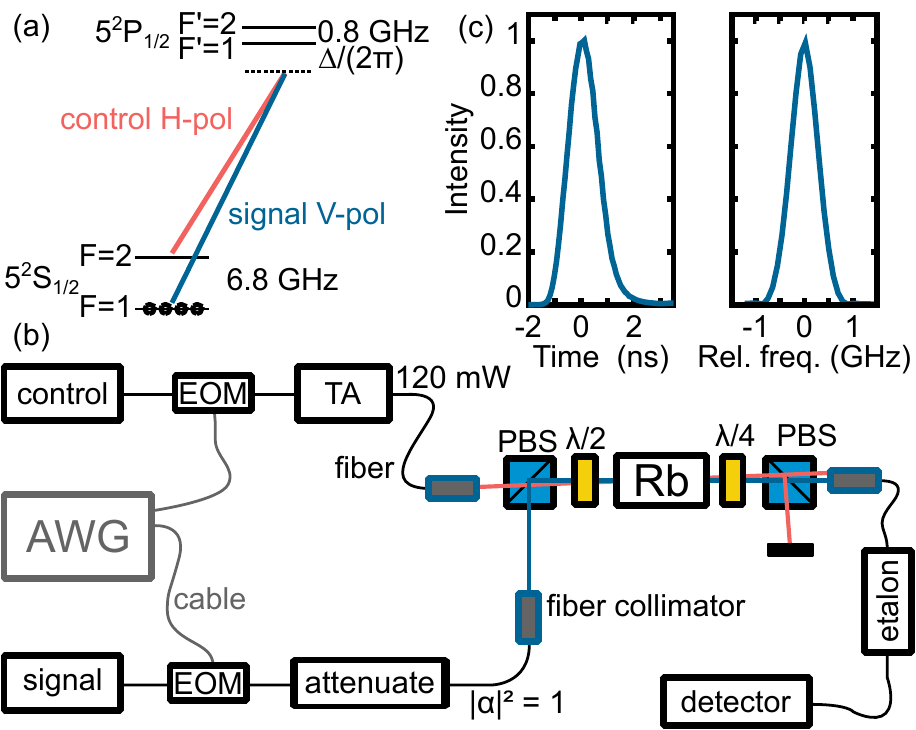}
\caption{ (a) Energy levels of the Rb $\textrm{D}_1$ line and transitions involved in the memory experiments. Virtually all atoms are initially prepared in the $F=1$ ground state. The  vertically  polarized signal to be stored is detuned by $\Delta$ from the $F=1\rightarrow F'=1$ transition, while the horizontally polarized control laser is detuned by  $\Delta$ from the $F=2\rightarrow F'=1$ transition. (b) Experimental setup for the memory experiment. EOM: electro-optic modulator; AWG: arbitrary waveform generator; TA: tapered amplifier;  PBS: polarizing beam splitter; Rb: vapor cell;  $\lambda/2, \lambda/4$: waveplates; detector: single avalanche photo diode (APD), a pair of APDs in a Hanbury Brown and Twiss configuration, or a single APD at one output of a heavily unbalanced Mach-Zehnder interferometer. (c) Shape of the signal pulses used in the storage and retrieval experiments measured with 250 ps timing resolution (left panel) and its Fourier transform (right panel). The latter indicates a FWHM bandwidth of 0.66~GHz.
}
\label{fig:1}
\end{figure}

One approach to tackle the bandwidth mismatch is to use a far-detuned Raman scheme and dense alkaline vapors \cite{Reim2010}. 
However, this scheme is intrinsically prone to four-wave-mixing (FWM) noise \cite{Michelberger2015}, impeding experiments in the quantum regime. Only very recently and with significant technological effort could the FWM problem be circumvented, in an experiment with a cavity-enhanced Raman memory \cite{Saunders2015}. 
While the achieved total internal efficiency was on the order of 10\%, the end-to-end memory efficiency was further reduced by about 3 orders of magnitude by the filtering system and impedance matching issues that are hard to avoid in cavity enhanced GHz bandwidth memories \cite{Afzelius2010}.

In contrast, we employ a cavity-free near-resonant memory scheme based on electromagnetically induced transparency (EIT) In principle, this scheme can achieve near unity end-to-end efficiencies and sufficient storage bandwidth, while being unaffected by FWM noise \cite{Rakher2013} and allowing for spatial multimode operation. 
For coherent photon wavepackets with FWHM bandwidth of 0.66~GHz  we achieve $\eta_{\textrm{e2e}}^{\textrm{50ns}} = 3.4(3)\%$ end-to-end efficiency of the fiber-coupled memory, outperforming  previous experiments on broadband storage by about two orders of magnitude~\cite{Saunders2015}, while still leaving plenty of room for future improvements. For a coherent pulse containing one photon on average, the achieved signal to noise ratio is SNR = 3.7(6), demonstrating that FWM noise is indeed suppressed in the EIT memory scheme.
We find that EIT-based vapor cell memories are surprisingly well suited for the technologically relevant $\delta f\approx1$~GHz bandwidth regime, where the storage and retrieval processes are faster than the decoherence rates of the atomic excited states~\cite{Manz2007}.
 
We implement the EIT-based memory on the Rb $\textrm{D}_1$ line at 795 nm, with the level scheme shown in Fig.~1(a). 
Initially, all atoms are prepared in the $F = 1$ hyperfine state of the $5^{2}S_{1/2}$ ground state manifold by optical pumping. The signal to be stored is generated by a laser red detuned by $\Delta=-2\pi\cdot0.9$~GHz from the $F = 1 \rightarrow F'=1$,
% transition of the Rb $\textrm{D}_1$ line
while the control laser is equally detuned by $\Delta$ from the $F = 2 \rightarrow F'=1$ transition.
In the experiment signal pulses with a 1~ns fall time from 90\% to 10\% signal level are generated by modulating an attenuated laser with a fiber-integrated electro-optic modulator, controlled by a fast arbitrary waveform generator, see Fig. 1(b,c). The rise time of the signal is limited to 500~ps by the  RF components. The numerical Fourier transform of the measured pulses indicates the FWHM bandwidth of 0.66~GHz. 
The generated pulses are comparable to the envelope of QD photons, in particular, but not only when photon shaping techniques are applied \cite{Beguin2017}. 
The laser intensity is carefully adjusted such that each pulse in the fiber going into the memory setup contains, on average, $|\alpha|^{2}=1.0(1)$ photon, where the error originates from the uncertainty of the detection efficiency of $\eta_{\textrm{APD}}=60(6)\%$ of the single photon counting avalanche photo diode (APD) used for calibration. To ensure a stable average photon number during the experiments, a constant fraction of the pulse intensity is continuously monitored on a dedicated APD.
Similarly, control pulses with Gaussian envelope (FWHM of 5 ns) are generated by modulating a second laser and subsequent amplification in a tapered amplifier (TA). 
Amplified spontaneous emission from the TA is suppressed by a combination of a narrow band interference filter (0.3 nm FWHM) and a monolithic etalon (finesse 50, 54 GHz free spectral range), allowing for a maximum control power of $120$ mW measured in continuous wave operation. For the Rb $\textrm{D}_1$ transition, with a dipole moment of $d=2.54\cdot10^{-29}$~C$\cdot$m, and the given $e^{-2}$ beam diameter of 525 $\mu$m, this corresponds to a Rabi frequency $\Omega\approx2\pi \cdot 600$~MHz.
The control and signal pulses are combined on a polarizing beam splitter and aligned to enclose an angle of $10(1)$ mrad. This angle is sufficiently small to allow for good overlap between the signal  ($e^{-2}$ diameter: 400 $\mu$m) and control  beams  over the entire vapor cell (length: 37.5 mm).
The vapor cell is filled with isotopically enriched $^{87}$Rb and 11~Torr (about 15~mbar) N$_{2}$ buffer gas to reduce radiation trapping and to increase the optical pumping efficiency~\cite{Thomas2016}. It is heated to 75$^\circ$C to achieve an optical depth of OD~$ = 5$ on the $\textrm{D}_1$, $F = 1 \rightarrow F'=1$ transition when virtually all atoms are prepared in the $F=1$ ground state.

A major challenge in all quantum memory experiments is the separation of the weak signal from the strong control laser. We apply a combination of polarization filtering, spatial filtering with a single mode fiber, and spectral filtering with a monolithic Fabry-Perot etalon (finesse 50, 27.2~GHz free spectral range). The filtering system suppresses the control beam by more than 12 orders of magnitude (120 dB), while the signal pulses are attenuated by only a factor of 3 (4.8 dB). Signal transmission is mainly limited by optical components without anti-reflection coating, the non-optimized transmission bandwidth of the etalon, and non-ideal mode matching when coupling into fibers, each accounting for about $1.25$ dB signal attenuation.
\begin{figure}
\center
\includegraphics[width=0.9\columnwidth]{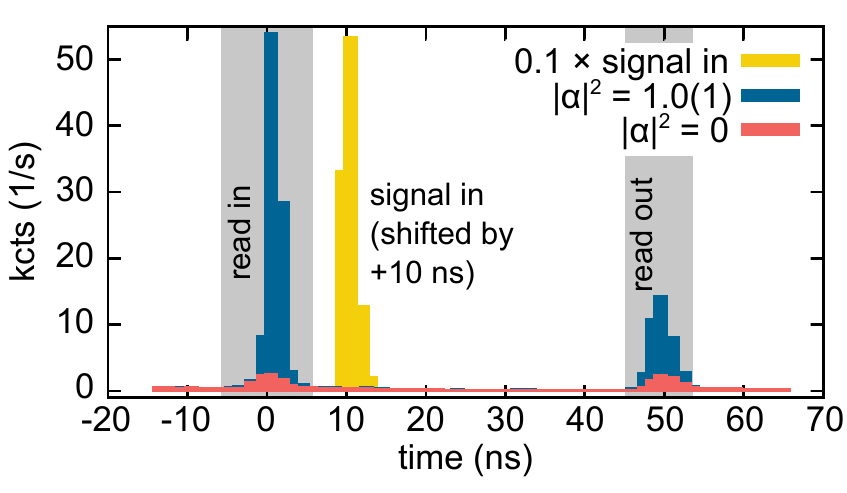}
\caption{Arrival time histogram for the photons detected in a memory experiment with a storage time of 50 ns, for a coherent input state with the envelope shown in Fig. 1(c) containing one photon on average [$|\alpha|^2=1.0(1)$], and for blocked input signal ($| \alpha |^2=0 $). Bin size is 1.3~ns. The time shifted  input pulse is shown for reference ($0.1 \times \,$signal in). The measured, noise corrected end-to-end efficiency of the memory setup including the filtering system is $\eta_{\textrm{e2e}}^{50\textrm{ns}} = 3.4(3)\%$, while the  
signal to noise ratio SNR~=~3.7(6) for the single photon level input pulse. }
\label{fig:2}
\end{figure}

In a storage and retrieval experiment, the memory is initialized by switching on the control laser for $500$~ns. After a subsequent waiting time of $25$~ns virtually all atoms have been optically pumped into the $F=1$ ground state and the signal and control pulses are sent into the vapor cell. After the storage time $T = 50$~ns, a second identical control pulse is applied for readout. Photons at the output fiber are detected with a single photon counting APD (Excelitas, timing resolution 350 ps) and a time-to-digital converter (qutools, timing resolution 81 ps). The experiment is repeated at a rate of $f_{\textrm{rep}} = 1.67$~MHz. Figure 2 shows an arrival time histogram of the photons detected during storage and retrieval. 
Within $t_{\textrm{int}} = 1$~s integration time $N_{\textrm{signal}}=42\cdot10^{3}$~photons are detected within the retrieval window. When the input signal is blocked, $N_{\textrm{noise}}=9\cdot10^{3}$~photons  are detected within the same retrieval window. From this, we infer the noise corrected  end-to-end efficiency 
$\eta_{\textrm{e2e}}^{50\textrm{ns}} = (N_{\textrm{signal}}- N_{\textrm{noise}})/(|\alpha|^{2} \,\eta_{\textrm{APD}} \,  f_{\textrm{rep}} t\,_{\textrm{int}})= 3.4(3)\%$.
 The signal to noise ratio for the retrieval of a single photon pulse is $\textrm{SNR } = (N_{\textrm{signal}}- N_{\textrm{noise}})/N_{\textrm{noise}}= 3.7(6)$, i.e. for $\mu_{1}=0.27(4)$ input photons the SNR would be unity~\cite{Jobez2015}.
 
 When correcting for the attenuation of the filtering system, we find an intrinsic memory efficiency of $\eta_{\textrm{int}}^{50\textrm{ns}} = 10(2)\%$ for storage and retrieval after $T=50$~ns.
The presented memory is not at all optimized for storage time and the $1/e$ memory lifetime is measured to be only $68(7)$ ns. The Gaussian decay of the retrieval signal indicates that atoms diffusing in and out of the narrow interaction volume is the main limitation for the storage time. Due to the high bandwidth of the memory, this comparably low value still allows for a time bandwidth product~\cite{Nunn2013} on the order of $B=100$. In future experiments, storage lifetime will be extended by several orders of magnitude by using larger beam diameters and magnetic shielding of the vapor cell. When taking the decay of the retrieval signal after 50 ns storage time into account, we find an total intrinsic memory efficiency of $\eta_{\textrm{int}} = 17(3)\%$.

\begin{figure}
\center
\includegraphics[width=0.9\columnwidth]{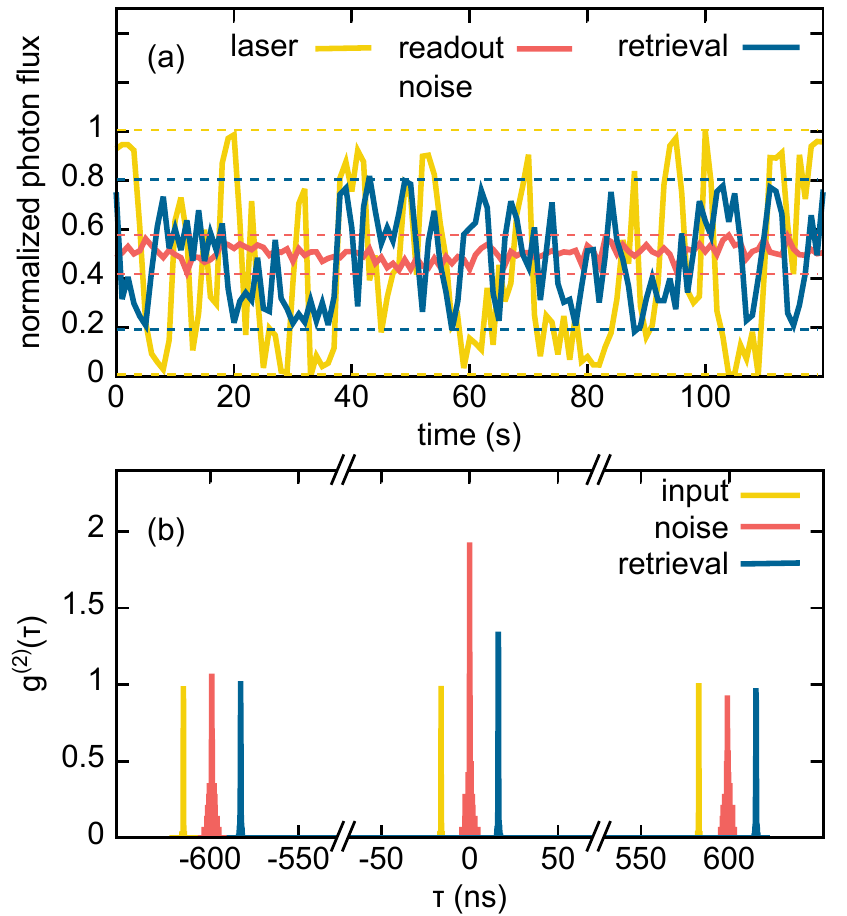}
\caption{(a) Typical interference between two subsequently stored and retrieved pulses, measured via thermal phase fluctuations of an unbalanced Mach-Zehnder interferometer. Integration time is 1~s for each data point. For a single photon level input pulse, a fringe visibility of $V=0.65(5)$ is achieved (curve ``retrieval''). The read-out noise shows a fringe visibility below $V=0.15$ originating from random intensity fluctuations.  To compensate for an imperfect mode overlap in the interferometer, data is normalized such that the input laser shows unity fringe visibility $V=\frac{max-min}{max+min}$, where $max$ ($min$) refers to the maximum (minimum) observed photon flux indicated by dashed lines.
(b) Second order autocorrelation $g^{(2)}(\tau)$ of photons detected during input pulse and read-out process, respectively. Due to contamination by noise, the readout signal shows a $g^{(2)}(0) = 1.3$, while  $g^{(2)}(0) = 1.0$ is expected in a perfect memory experiment. As expected for thermal (coherent) radiation, the readout noise (input signal) exhibits $g^{(2)}(0) = 2.0$ ($g^{(2)}(0) = 1.0$). Data is normalized to the peaks at $\pm600$~ns and input signal (retrieval) is shifted by -16~(+16)~ns for better visibility.}
\end{figure} 
To be suitable for quantum applications, it is important that the coherence properties of the light are preserved during the storage process~\cite{Kupchak2015}. 
In particular, it is necessary to preserve the mutual coherence between two subsequently stored and retrieved pulses, as well as the second order autocorrelation function of the retrieved signal~\cite{Sangouard2011}. For measuring coherence between two retrieved pulses, an unbalanced fiber-based Mach-Zehnder interferometer with $400$~ns arm length difference is inserted prior to photon detection. At the same time, the repetition rate  for the storage experiment is adjusted to 1/(400~ns) to make the retrieval signal interfere with its time shifted dublicate. 
Figure 3(a) shows a time trace of the photon flux measured at one interferometer output port, while the interferometer's phase given by the arm length difference randomly drifts, driven by thermal fluctuations. 
As compared to the reference signal generated by the input laser pulses, the fringe visibility is reduced to $V=0.65(5)$ for the stored and retrieved light.  When increasing the average photon number in the input pulses from $|\alpha|^{2}=1.0$ to $|\alpha|^{2}=10$, almost perfect interference of the retrieved signal with a visibility $V > 0.99(1)$ is achieved. This indicates that the reduced visibility in the single photon-level experiment mainly originates in the contamination of the signal by broadband read-out noise that shows no interference, where the SNR = 2 is decreased compared to the measurement shown in Fig. 2.  At higher signal intensity the influence of noise photons becomes negligible and consequently almost perfect interference is observed.
\begin{figure}
\center
\includegraphics[width=0.9\columnwidth]{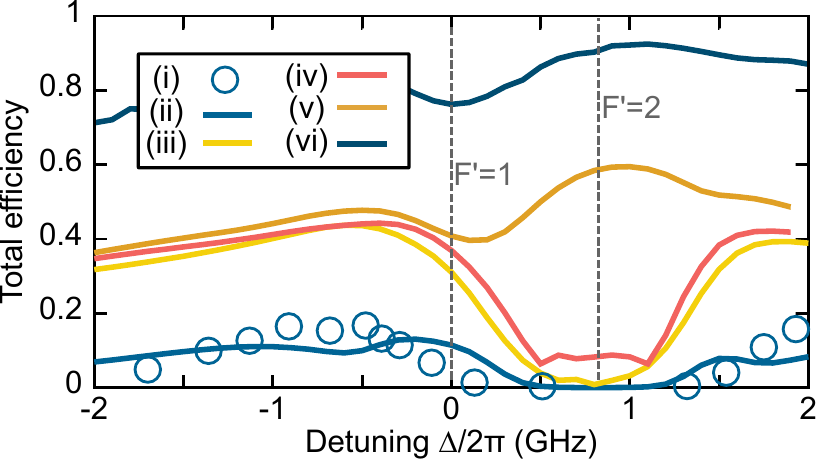}
\caption{ Total efficiency for storage and subsequent retrieval of the experimentally realized signal as a function of the detuning $\Delta$ for various situations. (i): Measured data. Simulation with same OD and level scheme as in the experiment for (ii): the experimentally realized Gaussian control pulses with 120~mW peak power, (iii): Gaussian control pulses with higher laser power,  (iv): optimal control pulses.  Simulation with suppressed parasitic single-photon transitions for (v): OD = 5 and (vi): OD = 35.}
\label{fig:4}
\end{figure}
The second-order autocorrelation function at zero time delay $g^{(2)}(0)$ of the retrieved photons, measured with a Hanbury Brown and Twiss setup (Fig. 3(b)), shows a slightly increased value of $g^{(2)}(0) = 1.3$, compared to the coherent input signal exhibiting $g^{(2)}(0) = 1.0$. Again, this is explained by noise contamination: When the input signal is blocked, $g^{(2)}(0) = 2.0$ is measured for the readout noise as expected for thermal light.  
Apart from contamination with broadband fluorescence, the high bandwidth  memory preserves coherence and statistical properties of the input state. 
This is in contrast to previous experiments on low bandwidth memories that are known to be prone to decoherence of the atomic excited state \cite{Manz2007}.  

To understand the limitations of the present memory experiment, we performed numerical simulations of the system along the lines of Ref. \cite{Gorshkov2008}, including Doppler broadening  of the atomic transitions by 500~MHz \cite{Gorshkov2007a}, and the doublet structure of the excited state \cite{Rakher2013}. By selection rules, the $m_{F'} = 0$ Zeeman sub-level of the $F'=2$ excited state  does not couple to the linearly polarized control laser~\cite{Yan2001}, which was modeled as a parasitic single-photon transition \cite{footnote1}. 
Dependent on the detuning $\Delta$, up to $\eta_{\textrm{int}}=16\%$ total internal efficiency is predicted for the experimentally realized situation, in excellent agreement with the measured data, see Fig.~4. Using Gaussian control pulses with 4 times higher peak Rabi frequency up to $\eta_{\textrm{int}}=43\%$ can be achieved, while optimal control pulses allow for $\eta_{\textrm{int}} = 45\%$. When 
unwanted excitations of the  $F'=2, m_{F'} = 0$ state is prevented $\eta_{\textrm{int}} = 60\%$ ($\eta_{\textrm{int}} = 92\%$) is predicted for OD = 5 (OD = 35).

In summary, we have demonstrated single-photon-level operation of a technologically simple atomic quantum memory for light pulses with a bandwidth of almost 1~GHz.
The pulse intensity and temporal envelopes are comparable to single photons emitted by state-of-the-art semiconductor quantum dot (QD) sources.
The memory will be directly suitable as node in a quantum network, if combined with such a source based on GaAs/AlGaAs QDs.  
Future work will be devoted to achieving storage and retrieval of true single photons emitted by a QD single photon source. The noise level will be reduced by enhancing the pumping efficiency into the $F = 1$ ground state over the entire vapor cell, e.g. by using a dedicated pump laser with rather large beam diameter and by using anti-relaxation coated cells. This will prevent atoms in the wrong ground state ($F = 2$) from diffusing into the interaction volume and generating readout noise. If necessary,  FWM noise can be suppressed by selectively preparing one Zeeman sub-level and exploiting selection rules. The storage time will be extended into the sub-ms regime by increasing the beam diameters (which define the interaction volume) and employing magnetic shielding to reduce ground state decoherence.  The memory efficiency will be boosted by optimizing the filtering system and using control pulses with higher power.  These straightforward improvements will boost the end-to-end-efficiency to approach $\eta_{\textrm{e2e}}=35\%$. Finally, suppressing the parasitic single-photon transitions by exploiting the Zeeman sub-structure and  selection rules and increasing the optical depth will push atomic vapor quantum memories for QD single photons towards unity efficiency, enabling truly scalable quantum networks.

The research leading to these results has received funding from the Marie Sk\l{}odowska-Curie Actions of the EU Horizon 2020 Framework Programme under grant agreement No. 702304 (3-5-FIRST) and from the Swiss National Science Foundation via NCCR Quantum Science and Technology.


\begin{thebibliography}{10}
\bibitem{footnote1}
Details on the simulations are outside the scope of this letter and will be published in a dedicated paper.
\end{thebibliography}
\end{document}